# Self-Similarity and Criticality in the Boson (pions) field produced according to dynamic cascade models.

Yiannis Contoyiannis[(1)] and Myron Kampitakis[(2)]

(1) Department of Electric-Electronics Engineering, West Attica University, 250 Thivon and P. Ralli, Aigaleo, Athens GR-12244, Greece (email: yiaconto@uniwa.gr)
(2) Hellenic Electricity Distribution Network Operator SA, Network Major Installations Department, 72 Athinon Ave., N.Faliro GR-18547, Greece (email: m.kampitakis@deddie.gr)

**Abstract:** In this work we reveal that the inside-outside dynamic model for the production of fermion-antifermion pair (or quark-antiquark in QCD), is an out of equilibrium process. Properties of a complex system as the self-similarity and the critical point of a continuous phase transition in equilibrium appear. In this work we show that this is accomplished imposing a type of mapping process through a self-composition operator on a $QED_2$ bosons (or pions) field.

## 1. Introduction

A great open issue in High Energy physics is the problem of confinement-deconfinement transition between the matter of Quarks-Gluons plasma and the Hadrons matter. This transition has cosmological significance since it refers to the production of the material of the Hadrons (baryons-mesons) at the early time of the universe. Today the problem is faced by two basic research currents. The first one supports that this transition was an out of equilibrium transition with clear dynamic characteristics while the second one suggests this transition has the characteristics of phase transition in equilibrium, as described by Statistical Physics of critical phenomena. The main argument for the out of equilibrium process is that the very small times in Hadronic interactions ($10^{-23}$s) did not permit the phase transitions mechanism and the main argument of the other side is that in critical state there is space-time invariance and therefore time scaling has no meaning. A possible scenario in the framework of the out of equilibrium process which has been developed in the past by Bjorken [1] in order to explain the mesons (quarks-antiquarks pair) production is the Inside-Outside cascade model. In the framework of the cascade models a scenario was introduced by Schwinger [2] which faces the particle production with QCD nonperturbative models. It is a two-dimensional quantum electrodynanics ($QED_2$) model in which the property of quark confinement is a natural consequence of dynamical process. On the other hand, various models have been developed in the framework of Statistical phase transitions mainly in numerical experiments on Lattice non-perturbative QCD [4]. In the level of symmetries, the SU(2) gauge group describes the two color (or flavors) degrees of freedom of QCD and has the same center with the Z(2) spin model as the Ising models. We have shown [3] that at critical point the fluctuations of the order parameter obey to a dynamic in time behavior which is an intermittent map. The motivation of this work is the idea that the dynamic mechanisms of bosons (pions) production like $QED_2$ can be simulated by an iterative map through a self-composition operator which produces the critical intermittent dynamics. Then self-similar structures in the space of bosons (pions) field as well as critical fluctuations in terms of a continuous phase transition emerge.

## 2. $QED_2$ model

Generally, according to this model, it has been found that $QED_2$ involving massless fermions with electromagnetic interaction is equivalent to a free boson field ϕ with a mass $m = \frac{e}{\sqrt{\pi}}$, where e is the coupling constant. This is analogous to the system of quarks and antiquarks in QCD in which the physical quanta are mesons, which are bound states consisting a quark and an antiquark. The physical boson in $QED_2$ can be considered as the analogue of the pion of QCD. Starting from Klein-Gordon equation one can obtain [4] the produced boson field :

$$\phi_{pro}(x^1, t) = \frac{-1}{\sqrt{\pi}} \int \frac{dp^1}{p^0} \sin(p^0 t - p^1 x^1) + \sqrt{\pi}[\theta(x^1 + t) - \theta(x^1 - t)] \quad (1)$$

where $x^1$ the spatial direction where the boson as bound state (fermion-antifermion) moves. The quantities $p^0$ and $p^1$ are the components of momentum $p^\mu, \mu = 0,1$. According to this picture, the produced boson field $\phi_{pro}$ starts to emerge at the point of the separation of the fermion-antifermion pair and spreads from the inside (small $x^1$) region to the outside ( large $x^1$) region, as time proceeds. In the inside-outside cascade picture it can be shown [1] that the produced boson field $\phi_{pro}$ is a function of the relativistic invariant quantity $= \sqrt{t^2 - (x^1)^2}$ , τ being the proper time. One can obtain the explicit proper-time dependence by considering $\phi_{pro}$ at the space-time point $(x^1 = 0, t)$. Thus we have from (1) :

$$\phi_{pro}(\tau) = \frac{-1}{\sqrt{\pi}} \int \frac{dp^1}{p^0} \sin(p^0 \tau) + \sqrt{\pi}[\theta(\tau) - \theta(-\tau)]$$

$$= -\sqrt{\pi} J_0(m\tau) + \sqrt{\pi} \quad (2)$$

where $J_0(m\tau)$ is the Bessel function of order 0. From equation (2) we conclude that the produced field $\phi_{pro}$ oscillates about $\sqrt{\pi}$. In the following we let m=1. The Bessel function is expressed through its series expansion:

$$J_0(\tau) = \sum_{k=0}^{\infty} \frac{(-1)^k (\frac{\tau}{2})^{2k}}{(k!)^2} = 1 + \sum_{k=1}^{\infty} \frac{(-1)^k (\frac{\tau}{2})^{2k}}{(k!)^2}$$

in eq.(2) obtaining that the bosons (or pions) Field $F(\tau) \equiv \frac{\phi_{pro}(\tau)}{\sqrt{\pi}}$ is given as :

$$F(\tau) = -\sum_{k=1}^{\infty} \frac{(-1)^k (\frac{\tau}{2})^{2k}}{(k!)^2} \quad (3)$$

Fig.1 illustrates the boson field $F(\tau)$ as a function of $\tau$.

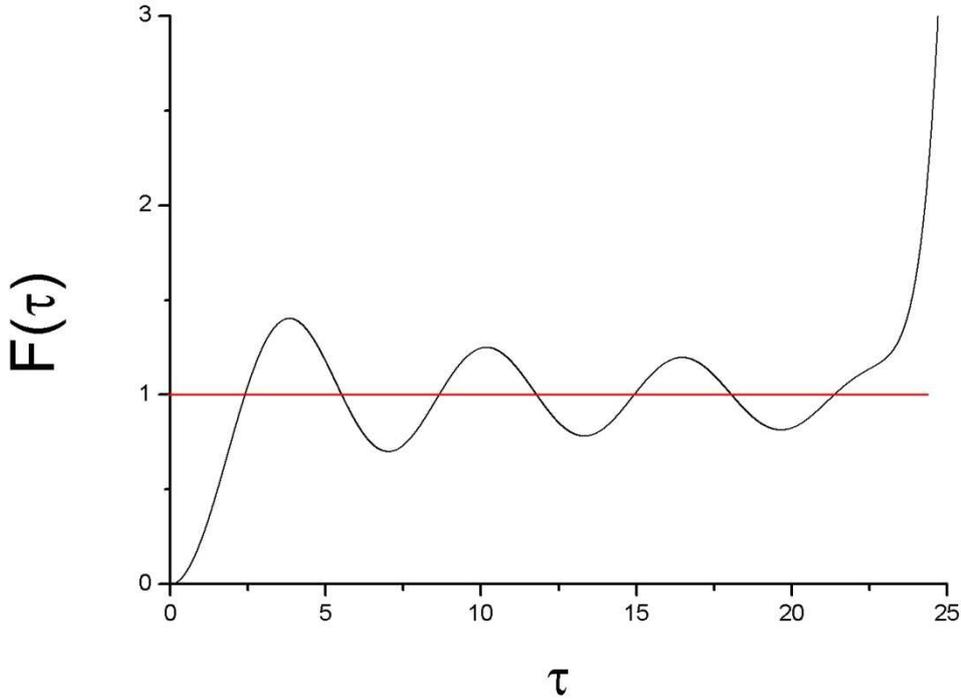

*Fig.1 The produced boson field $F(\tau)$ as a function of the proper time $\tau$. The points where the red line intersects the curve are the time locations where pair the fermion-antifermion (quark-antiquark for QCD) is created. The occurring non convergence for large values of $\tau$ is due to finite calculation capacity of the machine.*

# 3. A mapping process creates Self-Similarity on *boson field* $F(\tau)$

According to the inside-outside cascade process the $q\bar{q}$ pair after its production, it is separated. According to the mechanism described in [5] the interaction between a quark and an antiquark can be represented phenomenologically by a linear potential proportional to the separation between the two particles which is an essential feature for quark confinement. When the separation exceeds a threshold value, a new pair $q\bar{q}$ can be produced at a next instant in such a flux tube. This process continues producing pairs $q\bar{q}$ at successive time instants. We propose that the above iterative process could be mathematically expressed by the action of a self-composition operator. The composition function is an operation that takes two functions *f* and *g* and produces a function *h* such that h(x) = g(f(x)). If $g \equiv f$ then we can write $f^2(x) = f(f(x))$. For any natural number *n* ≥ 2, the *n*th functional power can be defined as f(f(f(....(f(x)...)) = $f^n$(x). The repeated composition of such a function with itself is called a self-composition function. Essentially, this process is a mapping of boson field $F(\tau)$ on to itself. In figure 2 the boson field $F^4(\tau)$ vs $\tau$ is presented.

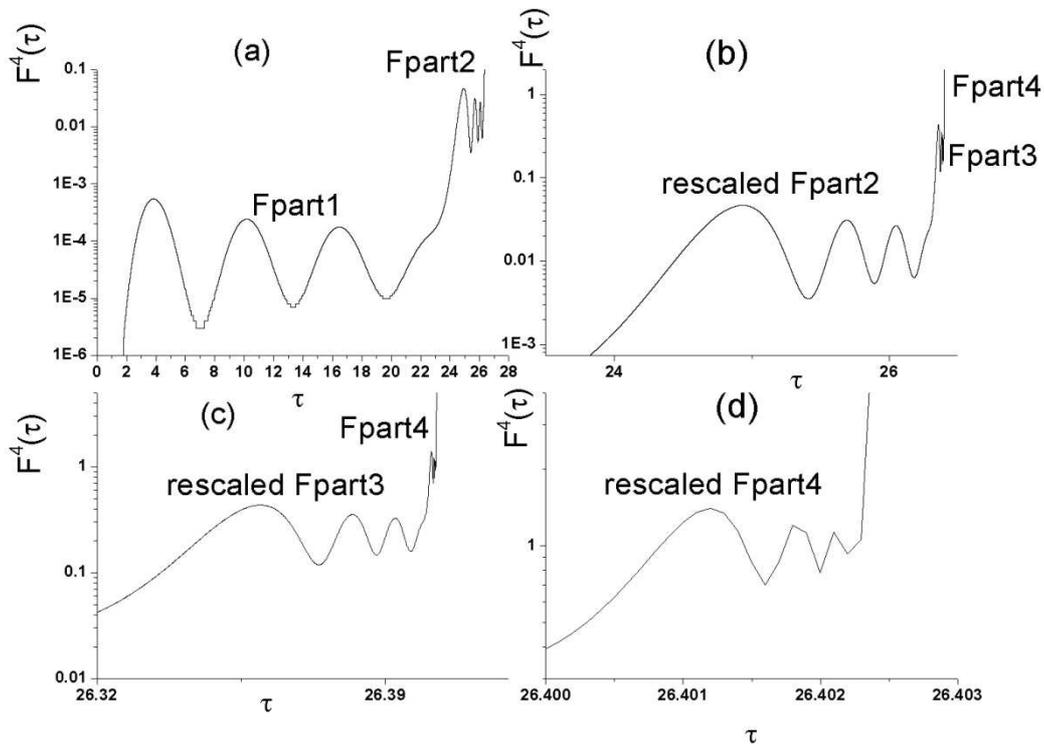

*Fig2. The 4$^{th}$ iteration field $F^4(\tau)$ vs $\tau$ is shown. The self-similarity between the parts of the boson field $F^4(\tau)$ is obvious.*

The existence of self-similar structures due to the action of the self-composition operator is apparent. The self-similar structures are characterized by scale invariance, such that if a part of a system is magnified until it is as large as the original system (figs 2(b),(c),(d)), one would not be able to tell the difference between the magnified part and the original system [6]. It is essential that the self-similarity here is not coming from any scale factor in the case of doubling operator Tf(x)=af(f(x/a)) which is met in renormalization group processes in maps [7].The generation of self-similar structures is the result of the action of self-composition operator only. One could say that this mapping shifts the limits of convergence of boson field towards greater times. Indeed, comparing fig 1 and fig2 it is clear that the radius of convergence has shifted from $\tau \approx 22$ to $\tau \approx 26,4$ where the fourth structure appears. As N increases, new self-similar structures could be added which belong to boson (pions) field $F^N(\tau)$. Scaling relations can be deduced between the self-similar structures corresponding to various values of number iterations. Such a scaling relation can be found for the envelope describing the decay of boson field. The curve function for the part1 of boson field for N=2,3,4,… is given by power-law

$$y(\tau) \sim \tau^{-d_N} \; ; \; d_N = \frac{2^{N-2}}{5} \quad (4)$$

Scaling law (4) can be applied to other parameters too.

## 4. Criticality in the Boson Field

### 4a. The method of critical fluctuations

The existence of self-similarity in the limit of convergence leads us to investigate if a critical state according to the Ginzburg –Landau Theory exists in this area because the self-similarity is a basic characteristic of critical state. It is very important in an out of equilibrium process of boson ( or pions for QCD) production as the inside-outside model is connected with a continuous phase transition in equilibrium following the theory of critical phenomena. A way to find the existence of the critical state is the analysis of order parameter fluctuations $\phi$ by the Method of Critical Fluctuations (MCF) . Details for this method are presented in [8]. Importantly, the exact dynamics at the critical point can be determined analytically for a large class of critical systems introducing the so-called critical map [3]. This map can be approximated as a intermittent map:

$$\phi_{n+1} = \phi_n + u\phi_n + \varepsilon_n \quad (5)$$

The shift parameter $\varepsilon_n$ introduces a non-universal stochastic noise which is necessary for the creation of ergodicity [9]. Each physical system has its characteristic "noise", which is expressed through the shift parameter $\varepsilon_n$. Notice, for thermal systems the exponent z is connected with the isothermal critical exponent $\delta$ as $z = \delta + 1$. The crucial observation in this approach is the fact that the distribution $P(l)$ of the suitable defined laminar lengths $l$ (waiting times in laminar region) [7] of the above mentioned intermittent map of Eq. (11) in the limit $\varepsilon_n \to 0$ is given by the power law [7]

$$P(l) \sim l^{-p_l} \tag{6}$$

where the exponent $p_l$ is connected with the exponent $z$ by $p_l = \dfrac{z}{z-1}$. Therefore the exponent $p_l$ is connected with the isothermal exponent $\delta$ by: $p_l = 1 + \dfrac{1}{\delta}$. The distribution of the laminar lengths of fluctuations is fitted by the function:

$$P(l) \sim l^{-p_2} e^{-p_3 l} \tag{7}$$

We focus on the exponents $p_2$ and $p_3$. If the exponent $p_3$ is zero, then, the exponent $p_2$ is equal to the exponent $p_l$. The relation $p_l = \dfrac{z}{z-1}$ suggests that the exponent $p_l$ (or $p_2$) should be greater than 1. On the other hand according to the theory of critical phenomena [6] the isothermal exponent $\delta$ is greater than 1. Therefore, as a result from $p_l = 1 + \dfrac{1}{\delta}$ we take $1 < p_l(p_2) < 2$. In conclusion, the critical profile of the temporal fluctuations is restored in the restrictions: $p_2 > 1$ and $p_3 \approx 0$. As the system removes from the critical state, the exponent $p_2$ decreases while simultaneously the exponent $p_3$ increases reinforcing, in this way, the exponential character of the laminar lengths distribution.

## 4b. Results from MCF analysis

The MCF is a method which requires the existence of the fluctuations. It is known that if a dynamic process has critical dynamics then the addition of properly quantity of noise for ergodicity purposes helps this criticality to reveal[10]. Recently we introduced a method where the above idea is applied [11]. According to this method for a given signal where the short-time range fluctuations are absent we normalize this signal in the interval [0,1] and then we add uniform noise according to the [10]. In this way, we create a new time-series in which fluctuations appear and finally we apply the MCF on this modified signal. In figure 3 we present the results from MCF analysis from the part2 of the boson field $F^4(\tau)$ ( fig2b).

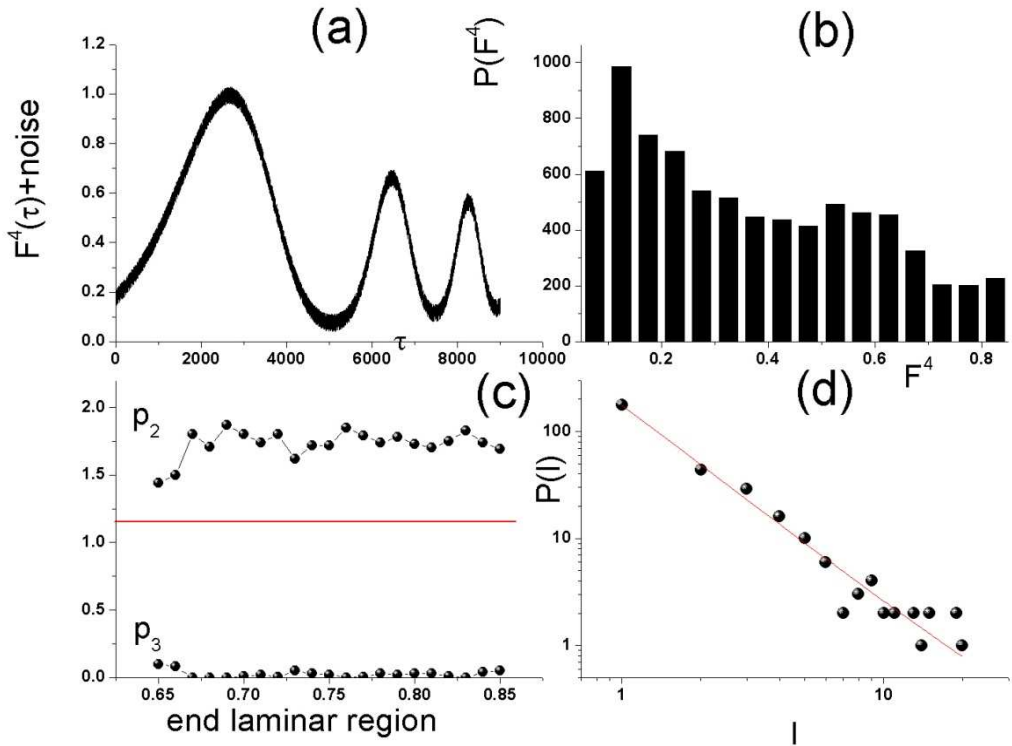

**Fig3**. *(a) The normalized structure of $F^4(\tau)$ in the interval [0.1] with noise addition. The noise is uniform in the interval [-0.035, 0.035]. The results do not change significantly for a broader region of noise between $\pm 0.03 - \pm 0.04$. (b) The histogram of $F^4(\tau)$ values. The fixed point (f.p) lies in the distribution edge with greater statistical weights because the f.p "attracts the trajectories". Here it has been set to 0.1. The ends of laminar regions lie in the area between 0.65-0.85 (c) The values of the exponents $p_2, p_3$ in the area of laminar regions vs the end of each laminar region are shown. It is characteristic that all the $p_2$ values are greater than 1 and simultaneously the $p_3$ is very close to 0. These are the critical conditions and so we can claim that critical fluctuations are present. The most representing $p_2$ values are these which correspond to $p_3$ exactly at the zero value because then the power-laws distributions appear. As we have seen the criticality is extended for $p_2 \in (1, 2)$. There are 5 such values fluctuating between 1.8 up to 1.87 which is close to the upper limits of criticality. (d) A representing laminar distribution fitted by the eq. (7) ( red line) for the laminar region $0.69 < F^4 < 0.1$. The fitting gives $p_2 = 1.87, p_3 = 0$ and $R^2 = 0.997$.*

The results of the above MCF analysis demonstrate clearly the existence of criticality in bosons (pions) field. Such a result appears in the critical point of 3D-Ising model [11] which as is known undergoes a continued phase transition in equilibrium.

# 6. Conclusions-discussion

Imposing the self-composition operator on the bosons( pions) produced field according to the cascade models mechanism, the self-similarity and criticality emerge in the bosons(pions) system in terms of Physics of critical phenomena. The existence of random noise is necessary to see the equivalence between the dynamic process and the criticality in stationary conditions. The decisive role for the emergence of criticality is the action of the self-composition operator. The crucial question is if every critical state in nature is a result of the action of self-composition operator on the field who plays the role of order parameter of a critical system. This is an issue for further investigation. In our case this happened since the cascade mechanism is essentially a feedback mechanism where the creation of a pair fermion-antifermion ( or quark-antiquark) acts as input in order to emerge the next pair . This is an iterative process which could mathematically be expressed by the self-composition operator as it is shown in this work .